\documentstyle[aps,preprint,epsf]{revtex} 
\newcommand{\be}{\begin{eqnarray}} 
\newcommand{\ee}{\end{eqnarray}}

\begin{document}
\title{A New Approach to y-scaling and the  Universal Features of Scaling Functions 
and  Nucleon Momentum Distributions} 
\author{Claudio Ciofi degli Atti} 
\address{ 
Dipartimento di Fisica, Universit\`a di Perugia and  Istituto Nazionale di 
Fisica Nucleare,\\ Sezione di Perugia Via A.Pascoli, I-06100 Perugia,\\ Italy } 
\author{Geoffrey B. West} 
\address{Theoretical Division, T-8, MS B285, Los 
Alamos National Laboratory, Los Alamos, NM 87545, USA} 
\date{\today} 
\maketitle

\begin{abstract}
 Some systematic general features of y-scaling structure functions, which  are essentially
 independent of detailed dynamics, are pointed out. Their physical interpretation in terms
of general characteristics, such as a mean field description and nucleon-nucleon
correlations, is given and their relationship to the momentum distributions 
illustrated. A new relativistic scaling variable is proposed which incorporates the momentum dependence of the excitation energy of the $(A-1)$ system,  with  the resulting scaling function being closely related to  the longitudinal momentum distributions and being free from removal-energy scaling violating effects. 
\end{abstract}
\newpage

Inclusive quasi-elastic electron scattering is potentially a powerful method for  
measuring the momentum distribution of nucleons inside a nucleus. Non-relativistically, this is most  
succinctly made manifest by expressing the data in terms of the scaling variable 
$y$ which, over a large kinematic range, can be identified as the longitudinal 
momentum  of the struck nucleon, $k_{\parallel}$ \cite{west}. At sufficiently 
large momentum transfers,  $\bf{q}$, the structure function, $W(\nu, q^{2})$, 
which represents the deviation of the cross-section from scattering from free 
nucleons, scales to a function of the single variable $y$ according to $qW(\nu , 
q^{2}) \approx f(y)$ where $\nu$ is the electron energy loss and $q\equiv 
\vert\bf{q}\vert$. Thus, in the scaling limit, $qW$ approaches a function that 
effectively traces out the longitudinal momentum distribution of the nucleons: 
\begin{equation} f(y) = \int n(k_{\parallel},{\bf{k}_{\perp}}) d^2 
{\bf{k_{\perp}}} =   2\pi\int\displaylimits_{\vert y\vert}^{\infty} n(k) k dk 
\label{one} \end{equation}  Here, $n(k)$ (with $k\equiv \vert\bf{k}\vert$) is 
the conventional nucleon momentum distribution function  normalized such that 
\begin{equation} \int{d^{3}k}n(k) = \int^{\infty}_{-\infty}dy f(y) = 1 
\label{three} \end{equation} Knowledge of $f(y)$ can therefore be used to obtain 
$n(k)$ by inverting Eq. (\ref{one}): \begin{equation} n(k) = - \frac{1}{2 \pi y} 
\frac{d f(y)} {dy} \hspace{1in}   \vert y\vert = k \label{four} \end{equation}

The above picture relies on the simple assumptions that: \begin{itemize} 
\item[i)] nucleon binding does not play a role in the scattering process; in 
reality, $qW(\nu, q^{2})$ is determined by the spectral function, $P(k,E)$, 
which depends both on the removal energy ($E$), as well as on the momentum of 
the nucleons, through the  relation \cite{ciofi} $qW(\nu , q^{2}) = F(y) = f(y) 
- B(y)$  where \begin{equation} B(y) = 2 \pi 
\int\displaylimits_{E_{min}}^{\infty}dE   \int\displaylimits _{\vert 
y\vert}^{k_{min}(y,E)}  {P_{1} (k,E)} \label{six} \end{equation} $P_{1}$ being  
that part of $P(k,E)$ generated by ground state correlations (thus, in a mean 
field description or, for the case of $^{2}H$, $P_{1}=0$)\cite{foot}; \item 
[ii)] final state interactions (FSI) are disregarded; if they are taken into 
account a direct relation between the experimentally measured $qW(\nu , q^{2})$ 
and the asymptotic scaling function $F(y)$ holds only approximately. 
\end{itemize} Over the past several years there have been vigorous theoretical 
and experimental efforts to explore $y$-scaling over a wide range of nuclei 
\cite{day}, using a relativistic scaling variable resulting from energy 
conservation implicit in the instant form of relativistic dynamics (see Eq. 
\ref{eleven} below) and representing the longitudinal momentum of those nucleons 
which have {\ the minimal value of $E$}  (recently, it has been shown \cite{poly} 
that, for the  deuteron, scaling in this variable  $y ( = y_r$ of ref. 
\cite{poly}) also follows from a relativistic light-front approach). 
In Ref. 
\cite{ciofi} the asymptotic scaling function $F(y)$ has been obtained by an 
elaborate extrapolation procedure of existing data
aimed at removing (or, at least, minimising) the effects of FSI.
The 
longitudinal  momentum distribution $f(y)$ has thereby been obtained by adding to $F(y)$ 
the binding correction $B(y)$ evaluated theoretically.
Such a procedure is based upon two basic assumptions: (i) the FSI can be represented as a power series in $1/q$, and die  out at large $q^2$, a conclusion which has been reached by various authors \cite{rinawest}
;(ii) the theoretical binding correction has to be applied to obtain $f(y)$ \cite {ciofi}. Both assumptions, which in principle could be questioned,  affect the final form of $f(y)$,  and therefore need to be investigated.
We will   discuss the effects  of FSI in  a separate paper \cite{faralli}; here we address only point (ii). To begin with, let us assume that  $f(y)$ obtained in \cite{ciofi}
is correct and let us analyze it in detail to see whether it contradicts or agrees with current theoretical predictions.
A systematic analysis, to be presented elsewhere \cite{faralli}, 
exhibits the following general features of $f(y)$ for nuclei with $A < 56$:

\begin{itemize}

\item[i)] $f(0)$ decreases monotically with $A$, from  $\sim 10^{-2} MeV^{-1}$ 
when $A=2$ to $\sim 3{\times 10^{-3}} MeV^{-1}$ for $^{56}Fe$; moreover, for $y 
\sim 0$, $f(y) \sim C_1(\alpha^{2} + y^{2})^{-1}$, with $\alpha$ ($C_1$) ranging 
from $\sim 45 MeV$ ($18 MeV$) for $A=2$, to $\sim 140 MeV$ ($59 MeV$) for 
$A=56$. \item[ii)] For $50MeV \le  \vert y\vert \le 200 MeV$,  $F(y) \sim 
e^{-a^{2}y^{2}}$ with $a$  ranging from $\sim 56 {\times 10^{-2}}MeV^{-1}$ for 
$A=2$, to $\sim 45 {\times  10^{-2}}MeV^{-1}$  for $A=56$. \item[iii)] For 
$\vert y\vert \ge 400 MeV$,  $f(y) \sim {C_2} e^{-b \vert y \vert}$, with $C_2$ 
ranging from $2.5 \times 10^{-4} MeV ^{-1}$ for $A=2$, to $6 \times 10^{-4} MeV 
^{-1}$ for $A=56$, and, most intriguingly, $b = 6 \times 10^{-3} MeV ^{-1}$,  
\it{independent} of $A$. \end{itemize}

The following  form for $f(y)$ yields an excellent representation of these 
general features for all nuclei: 
\begin{equation} 
f(y) = 
\frac{{C_1}e^{-a^{2}y^{2}}}{ \alpha^{2} + y^{2}}  + {C_2} e^{-b \vert y \vert} 
\label{seven}
\end{equation} 
The first term ($\equiv f_0$) dominates the small 
$y$-behavior, whereas the second term ($\equiv f_1$) dominates large $y$. The 
systematics of the first term are determined by the small and intermediate  
momentum behaviours of the single particle wave function. For $\vert 
y\vert\le\alpha$ this can be straightforwardly  understood in terms of a zero 
range approximation and is, therefore,  insensitive to details of the 
microscopic dynamics, or of a specific model. The small $k$ behavior of the 
single particle wave function is controlled by its separation energy, $(Q\equiv 
M + M_{A-1} - M_A = E_{min})$ and is given by  
$(k^{2}+\alpha^{2})^{-1}$ so  $\alpha = (2\mu Q)^{1\over 2}$, $\mu$ being the 
reduced mass of the nucleon.  

Before discussing the intermediate range it is instructive to consider first the 
large $y$-behavior. Perhaps the most intriguing phenomenological characteristic of
the data is that  {\it $f(y)$ falls off exponentially at large $y$ with a similar
slope parameter for all  nuclei, including the deuteron}. Since (i) $b$ is almost the same for all nuclei 
{\it including} $A=2$, i.e.,  $f(y)$, at large $y$, appears to be simply the rescaled 
 scaling function of the deuteron; and (ii)  
$b (\approx 1.18 fm) \ll 1/\alpha_D (\approx 4.35 fm)$, we conclude that the term
${C_2}e^{-b\vert y \vert}$ is related to 
the short range part of the deuteron wave function and reflects the universal nature 
of $NN$ correlations in nuclei.
The remaining parameters, $C_1$ and $a$, can be related to $f(0)$ and the 
normalization condition, Eq.~(\ref{three}). Once this is done, there are no 
adjustable parameters for different nuclei. The intermediate range is clearly 
sensitive to $a$, with the gaussian form being dictated by the shell model harmonic 
oscillator potential, modulated, however,  by 
the correct $\vert y\vert <\alpha$ behaviour, namely $(y^2 + \alpha^2)^{-1}$, 
thereby ensuring the correct asymptotic wave function. As an example, Fig.~1 
shows the  longitudinal scaling functions  $f(y)$  for $^2H$ and $^4He$ 
extracted from the experimental data  (\cite{ciofi}), compared to 
Eq.~(\ref{seven}). The errors of $f(y)$ for $^2H$ are very small, whereas, at large values of $|y|$  they are appreciable for $^4He$ and heavier nuclei; they    are due both to the lack of reliable experimental data at large values of $q^2$, and to the necessity to correct the asymptotic scaling function for binding effects, and produce an error on $C_2$ and $b$ in Eq.~(\ref{seven}), ranging from ${\simeq 6}$  $\%$  to ${\simeq 10}$  $\%$. A systematic analysis for a large body of nuclei 
exhibiting the same features as those shown in Fig.~1 will be presented 
elsewhere \cite{faralli}.

With these observations it is now possible to understand the normalization and 
evolution of $f(y)$ with $A$. First note that Eq.~(\ref{one}) implies $f(0) = 
{1\over 2}\int {d^3 k} {n(k)\over k} =  \langle 1/2k \rangle$ and so is mainly 
sensitive to small momenta. Now, typical mean momenta vary from around 50 MeV 
for the deuteron up to almost 300 MeV for nuclear matter.  We can, therefore, 
immediately see why $f(0)$ varies from around 10 for the deuteron to around 2-3 
for heavy nuclei. More specifically, since ${C_2}\ll {C_1}/\alpha^2$ and $f_1$ 
falls off so  rapidly with $y$, the normalization integral, Eq.~(\ref{three}), 
is dominated by small $y$, i.e., by $f_0$. This leads to $f(0) \approx 
(\pi^{1/2} \alpha)^{-1} = (2\pi\mu Q)^{-1/2}$ which gives an excellent fit to 
the $A$-dependence of $f(0)$. Since $f(y)$ is constrained by the sum rule, 
Eq.~(\ref{three}), whose normalization is independent of the nucleus, a decrease 
in $f(0)$ as one changes the nucleus must be compensated for by a spreading of 
the curve for larger values of $y$.  {\it Thus, an understanding of $f(y)$ for 
small $y$ coupled with an approximately universal fall-off for large $y$, 
together with the constraint of the sum rule, leads to an almost 
model-independent understanding of the gross features of the data for all 
nuclei}.

To sum up, the ``experimental" longitudinal momentum distribution can be thought 
of as the incoherent sum of a mean field shell-model contribution, $(f_{0})$, 
with the correct model-independent small $y$-behaviour built in, and a 
``universal" deuteron-like correlation  contribution $(f_{1})$. Thus, the 
momentum distribution, $n(k)$, which is obtained from (\ref{four}), is also a 
sum of two contributions: $n=n_{0} + n_{1}$. This allows a comparison with 
results from many body calculations in which $n_{0}$ and $n_{1}$ have been 
separately calculated. Of particular relevance are not only the shapes of 
$n_{0}$ and $n_{1}$, but also their normalizations, $S_{0(1)}\equiv\int 
n_{0(1)}d^3k = \int f_{0(1)}dy$ which, theoretically, turn out to be, for 
$^4He$, $S_{0} \sim 0.8$ and $S_{1} \sim 0.2$~\cite{panda} whereas 
Eq.~(\ref{seven}) yields $S_0 = 0.76$ and $S_1 = 0.24$. A comparison between the 
momentum distributions obtained from $y$-scaling and the theoretical ones is 
shown in Fig. 2. As can be seen the $n(k)$ compare very well with theoretical 
calculations. 
Our analysis confirms   very well known properties of low energy nuclear 
physics (binding, etc), as well as some features of the momentum distributions
predicted recently  by various theoretical calculations \cite{panda}, \cite{frank}, \cite{zabocps}, but still waiting for a firm experimental confirmation. To place our results on a more solid basis, it would be necessary however to reduce the errors on $f(y)$ at large values of $|y|$, which necessitates experimental data at larger values of $q^2$, as well as a reduction of the uncertainties related to the binding correction. Experimental data at higher $Q^2$ became recently available \cite{arrington}, and their inclusion in the analysis reduces indeed  the errors on $f(y)$ \cite{faralli}. Here we address the problem of making the extraction of $f(y)$ as much  independent as possible from theoretical binding corrections. This  is accomplished by introducing another scaling variable. A generic ambiguity in defining a scaling variable is that there is no  unique prescription for $y$, so that it is legitimate, in principle, to incorporate  in the definition of $y$ some  physical
dynamical effects by introducing proper effective parameters.
The usual scaling variable  $y$ is obtained 
from relativistic energy conservation 
\begin{equation} \nu + M_{A} = [(M_{A-1} + 
E_{A-1}^{*})^{2} + {\bf k}^2]^{1/2} + [M^{2} + ({\bf k} + {\bf 
q})^{2}]^{1/2} 
\label{eleven}
 \end{equation} 
by setting $k=y$, $\frac{{\bf 
k}\cdot {\bf q}}{kq} = 1$, and, most importantly, the excitation energy, 
$E_{A-1}^{*}=0$; thus, $y$ represents the nucleon longitudinal momentum of a 
nucleon having the {\it minimum value of the removal energy } $(E=E_{min}, 
 E_{A-1}^{*} = 0)$. The minimum value of the nucleon momentum when $q\to\infty$, 
becomes $k_{min}(y,E)=\vert y - (E - E_{min})\vert$. Only when $E=E_{min}$ does 
$k_{min}(y, E) = \vert y \vert$, in which case the binding correction $B = 0$ 
and $F(y)=f(y)$. However, the final spectator $(A-1)$ system can be left in all 
possible excited states,  including the continuum, so, in general, $E_{A-1}^{*} 
\ne 0$ and $E>E_{min}$, so $B(y) \ne 0$, and $F(y) \ne f(y)$. Thus, it is the 
dependence of $k_{min}$  on $E_{A-1}^{*}$ that gives rise to the binding effect, 
i.e. to the relation  $F(y) \ne f(y)$. This is an unavoidable defect of the 
usual approach to scaling; as a matter of fact, the longitudinal momentum is 
very different for weakly bound, shell model nucleons (for which ${E_{A-1}^*} 
\sim 0-20 MeV$) and  strongly bound, correlated nucleons (for which  
${E_{A-1}^*} \sim 50-200 MeV$), so that at large values of $|y|$ the scaling 
function is not related to the longitunal momentum of those nucleons (the 
strongly bound, correlated ones) whose contributions almost entirely exhaust the 
behaviour of the scaling function. In order to establish a global link between 
experimental data  and longitudinal momentum components, one has to conceive a 
scaling variable which could equally well represent longitudinal momenta of both 
weakly bound and strongly bound nucleons.
An attempt in such a direction has been made in the past by \cite{ji} and, recently,
by us \cite{nucl-th} with some minor differences with respect to Ref. \cite{ji} attempt. This was based upon taking literally  the two-nucleon correlation model according to which
 the large $k$ and $E$ behaviours of  the Spectral Function are governed by 
configurations in which the high momentum of a correlated nucleon (1, say) is 
almost entirely balanced  by another nucleon (2, say). Within such a picture, 
one obiously has
$E_{A-1}^{*} = \frac{A-2}{A-1}{\frac{1}{2M}}{\bf {k}}^2$ 
and the average excitation energy for 
a given value of $k$ is $<{E_{A-1}^*}(k)>=\frac {A-2}{A-1}\frac {{\bf 
k}^{2}}{2M}$. By replacing $E_{A-1}^*$ in  Eq.~(\ref{eleven}) with $<{E_{A-1}^*}(k)>$, the  deuteron-like scaling variable $y_{2}$ 
introduced in our previous paper \cite{nucl-th}(see also \cite{ji}, where a 
similar scaling variable was first introduced) is obtained, representing the 
scaling variable pertaining to a ``deuteron" with mass  $\tilde 
M=2M-E_{th}^{(2)}$, where $E_{th}^{(2)} = M_{A-2} +2M - M_{A}$. Such a scaling 
variable, however, has the unpleasant feature that the effect of the deuteron- 
like correlations are overestimated at low values of $y_2$ and , as a result, 
the correct shell-model picture provided by the usual variable $y$ is lost.

 In a more refined model, the CM motion of the pair is taken into account:
one has \cite{clasi} 
\begin{equation} 
E_{A-1}^{*} = \frac{A-2}{A-1}{\frac{1}{2M}}[{\bf {k}} -\frac 
{A-1}{A-2}{\bf {K}}_{CM}]^2 
\label{twelve}
 \end{equation} which shows that the 
excitation energy of the residual nucleus depends both upon $\bf k$ and ${\bf 
K}_{CM}$; although using Eq.~(\ref{twelve}) the situation is improved, the description ot low values of $y$ is not a satisfactory one.
In this paper we {\it assume} that the nucleus is described by a realistic spectral function as provided by few- and many-body calculations \cite {fewbody}, \cite{manybody} and implement such a realistic description into the definition of the scaling function. 
If the expectation value of Eq.~(\ref{twelve})  
is evaluated with realistic spectral functions, 
for the three-body system
\cite{fewbody} and nuclear matter \cite{manybody}
and the model spectral function of \cite{clasi}
one obtaines (\cite{faralli,clasi})
 \begin{equation} 
<E_{A-1}^{*}(k)> = \frac{A-2}{A-1}{\frac{1}{2M}}{\bf {k}}^2 +b_A-{c_A}{\frac{1}{2M}}{\bf 
k^{2}}. 
\label{tredici}
 \end{equation}
Here, $b_A$ and $c_A$,  resulting from 
the $CM$ motion of the pair, have values ranging  from $17 MeV$ to $43 MeV$ 
and $3.41 \times 10^{-1}$ to $1.66 \times 10^{-1}$, for $^3He$ and Nuclear 
Matter, respectively. Placing Eq.(\ref{tredici}) in Eq.(\ref{eleven}) and 
subtracting the value  of the average removal energy $<E>$ to counterbalance the 
value  (\ref{tredici}) at low values of $y$, a new scaling variable is obtained.
This effectively takes into account the $k$-dependence of the excitation energy 
of the residual $(A-1)$ system, both at low and high values of $y$. This is in contrast to the 
usual scaling variable, which completely disregards  $E_{A-1}^*$, and the 
scaling variable $y_2$, which overestimate the effects of deuteron-like 
correlations at low values of $y$. In the kinematical region of existing 
experimental data this new , $\it global$ scaling variable, $y_{CW}$,  that we have 
obtained,  reads as follows 
\begin{equation} 
y_{CW}={\bigg\vert} -{{\tilde 
q}\over 2} + \left [{{\tilde q}^2\over{4}} - \frac{4  {\nu_{A}}^2 M^{2}-\rm{ 
W_A}^{4}}{4\rm{ W_A}^{2}}\right ]^{1/2}{\bigg\vert} \label{quattordici} 
\end{equation} 
Here, $\nu_A = \nu + \tilde M$, $\tilde M = (2A-3)M/(A-1) - 
E_{th}^{(2)} -(b_A+2M^{2}c_A - <E>)$, $\tilde q = q -c_A{\nu}_A$ and 
$\rm{W}_{A}^{2} = {\nu_A}^2 - {\vec q}^2 =\tilde M^{2} + 2 \nu  \tilde M - 
Q^{2}$. For the deuteron $E_{A-1}^{*}=0$, so $y_{CW}\rightarrow y = \vert -q/2 + 
[q^{2}/4 - (4 {\nu_d}^{2} M^{2} - \rm{W_d}^{4})/{{\rm W_d}^{2}}]^{1/2}\vert$  
with ${\nu_d} = \nu + M_{d}$ and ${\rm W}_{d}^{2} =  {\nu_d}^2 - {\vec q}^2 =
{M_d}^2 + 2{\nu}M_d - Q^2$.  For small values of $y_{CW}$, such that 
${(\frac{A-2}{A-1}{\frac{1}{2M}}{y}^2 +b_A-{c_A}\frac{{y}^{2}}{2M})} \ll <E>$,  
the usual variable, representing the longitudinal momentum of a weakly bound 
nucleon is recovered \cite{foot4}. Thus $y_{CW}$ interpolates between the 
correlation and the single particle regions. More importantly, however, since 
$k_{min}(q,\nu,E) \simeq \vert y_{CW} \vert$, $B(y_{CW}) \simeq 0$,  $F(y_{CW}) \simeq 
f(y_{CW})$. 
One would therefore expect from our above  analysis, 
the same behaviour of $f(y_{CW})$ at high  values of $y_{CW}$ for both the deuteron 
and complex nuclei (unlike what happens with the usual scaling function $F(y)$), 
and the same shell-model behaviour at low values of $y$, as predicted by the 
usual scaling variable. This is, indeed, the case, as exhibited in Figs.~3 and 
4, where the direct, global, and independent of $A$ link between the  scaling 
function $F(q,y_{CW})$ and the longitudinal momentum distributions is manifest . 

We can summarise our conclusions as follows: \begin{itemize} \item[i)] The 
general universal features of the $y$-scaling function have been identified and 
interpreted in terms of  a model-independent zero-range 
contribution, a ``universal" 2-nucleon correlation contribution and a mean field 
(shell-model) contribution. The shape and evolution of the curve have 
been understood both  quantitatively and qualitatively on general grounds. 
\item[ii)] A global relativistic scaling variable which, unlike all previously proposed variables, incorporates the average mean field excitation energy  of the $(A-1)$ system, as well as the excitation energy produced by deuteron-like correlated pairs and by their center-of-mass motion in the nucleus, has been defined. Such a variable, thanks to these new features, allows one to establish a 
more direct link between the
scaling function and the longitudinal momentum distributions at any value of $y$. Thus, using this variable, it 
would be possible in principle
to obtain the  longitudinal momentum distributions  
directly   without introducing theoretical
binding 
corrections. 
Of course the usual variable has the advantage of being defined in terms of a well-defined experimental quantity, the minimum value of the removal energy $E_{min}$, whereas $y_{CW}$ incorporates the excitation 
energy of $(A-1)$ by a theoretical prediction. However, given the fact that 
in ${\gamma}^* - Nucleus$ scattering the virtual photon couples to 
nucleons having different values of the removal energy,  suggests that the removal energy has to be taken into account in the definition of the scaling variable.  Our Fig. 3 shows indeed that removal energy effects are very important. Since these are a source of scaling violation, the other source being the FSI, it seems reasonable to incorporate the binding effects into the definition of $y$ so as to ascribe the remaining scaling violation to FSI. Moreover the plot of the data in terms of $y_{CW}$ has a very clear cut meaning: $F(y_{CW},q^2)$ represents the scaling function at a given $q^2$ and for a longitudinal momentum of a nucleon having removal energy
$E_(y_{CW}) = E_{th}^{(2)} +\frac{A-2}{A-1}{\frac{1}{2M}}{y_{CW}}^2 +b_A-{c_A}{\frac{1}{2M}}
{ y_{CW}^{2}}$. As already pointed out, there is no unique prescription for defining a y-scaling variable, and various variables are currently being used \cite{variables}. Ultimately, a valid criterion for a scaling variable, is to produce scaling; the usual $y$ has been shown to produce scaling but, at large values of $|y|$, in the ultra  asymptotic limit \cite{ciofi}, \cite{gloeckle}, \cite{poly}, that is a limit which is not reached by present experimental data. In Ref. \cite{faralli}, it is shown that the new scaling variable not only produces precocious scaling to the longitudinal momentum distribution even at the largest value of $|y|$ recently reached by the new TJLAB experimental data \cite{arrington}, but also that the FSI on the scaling function $F(y_{CW},q^2)$ are very similar to the ones acting in the deuteron.  Therefore, in terms of this variable the data seem to support the idea that 
the large $y_{CW}$ behaviour in all nuclei is essentially nothing but a rescaled 
version of the deuteron (including, perhaps, also some effects from deuteron-like FSI, due to the constant value of the nucleon-nucleon cross section in the region ${2 \leq Q^2 \leq 6}$  $GeV^2$, as stressed in \cite{ji1}).  Such a conclusion cannot be reached by analysing the data in terms of the old scaling variable, which mixes up scaling violation due to removal energy effects and FSI.

We would like to thank Avraham Rinat and Dino Faralli for  useful comments and discussions.
 
\end{itemize}


\newpage

\begin{figure} 
\caption{
The "experimental" longitudinal momentum distributions 
of $^2H$ and $^4He$ obtained
\protect\cite{faralli} 
using the results of \protect\cite{ciofi} compared
with Eq. (\protect\ref{seven}) with $\alpha = 45 MeV$, $C_1 = 18 MeV$, $a = 56 \times 10^{-2} MeV^{-1}$, $C_2 = 2.5 \times 10^{-4} MeV^{-1}$, and  $b = 6   \times 10^{-3} MeV^{-1}$, 
 for $^2H$, and  $\alpha = 167 MeV$, $C_1 = 106 MeV$, $a = 68.5\times 10^{-2} MeV^{-1}$,
$C_2 = (6 \pm 0.6)\times 10^{-4} MeV^{-1}$, and  $b = (6 \pm 0.6) \times 10^{-3} MeV^{-1}$,
for $^4He$}.
\label{fig1}
\end{figure}

\begin {figure} 
\caption{The nucleon momentum distribution for $^2H$
from Eq.(\ref{eleven}), compared with the one obtained from the AV14 interaction;
the same as Fig. 2a but for $^4He$}
\label{fig2}
\end{figure}

\begin {figure} 
\caption{  The experimental scaling function $F(q,y)$ for $^{2}H$, $^{4}H$
and $^{56}Fe$ compared with the longitudinal momentum distributions $f(y)$ given by Eq.(\ref{seven}) (dot-dash-$^{2}H$; short-dash-$^{4}H$; full-$^{56}Fe$). The scaling variable is the usual one, i.e. the one obtained from Eq.(\ref{seven}) placing $E_{A-1}^* = 0$}
\label{fig3}
\end{figure}

\begin {figure} 
\caption{  The same as in Fig.3, but with the variable , $y_{CW}$, obtained using  Eq.(\ref{tredici})}
\label{fig4}
\end{figure}

\newpage
\epsfxsize 14cm
\centerline{\epsfbox{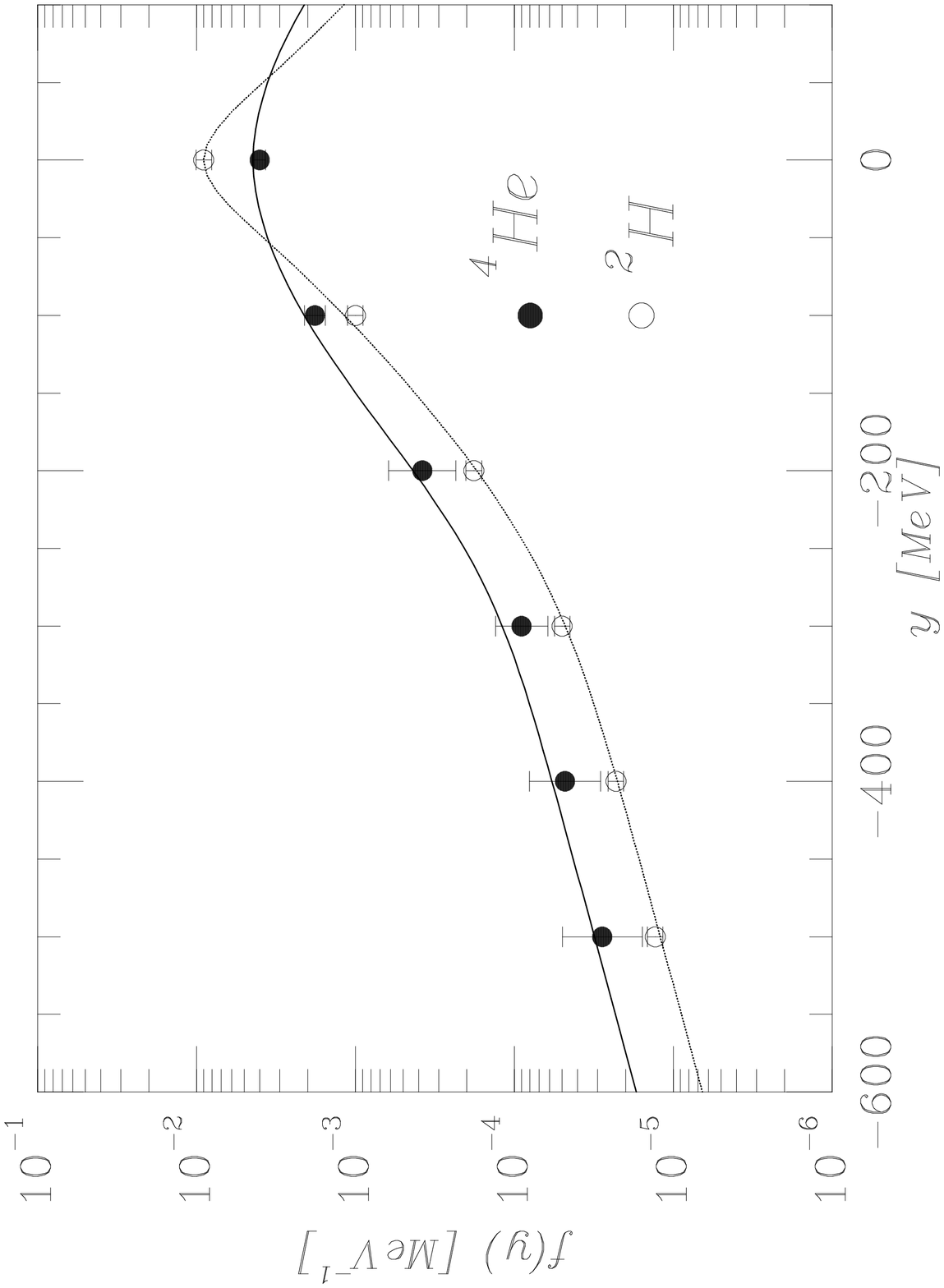}}
\vfill
Fig.~\ref{fig1}.
Ciofi-West Y-scaling

\newpage
\epsfxsize 14cm
\centerline{\epsfbox{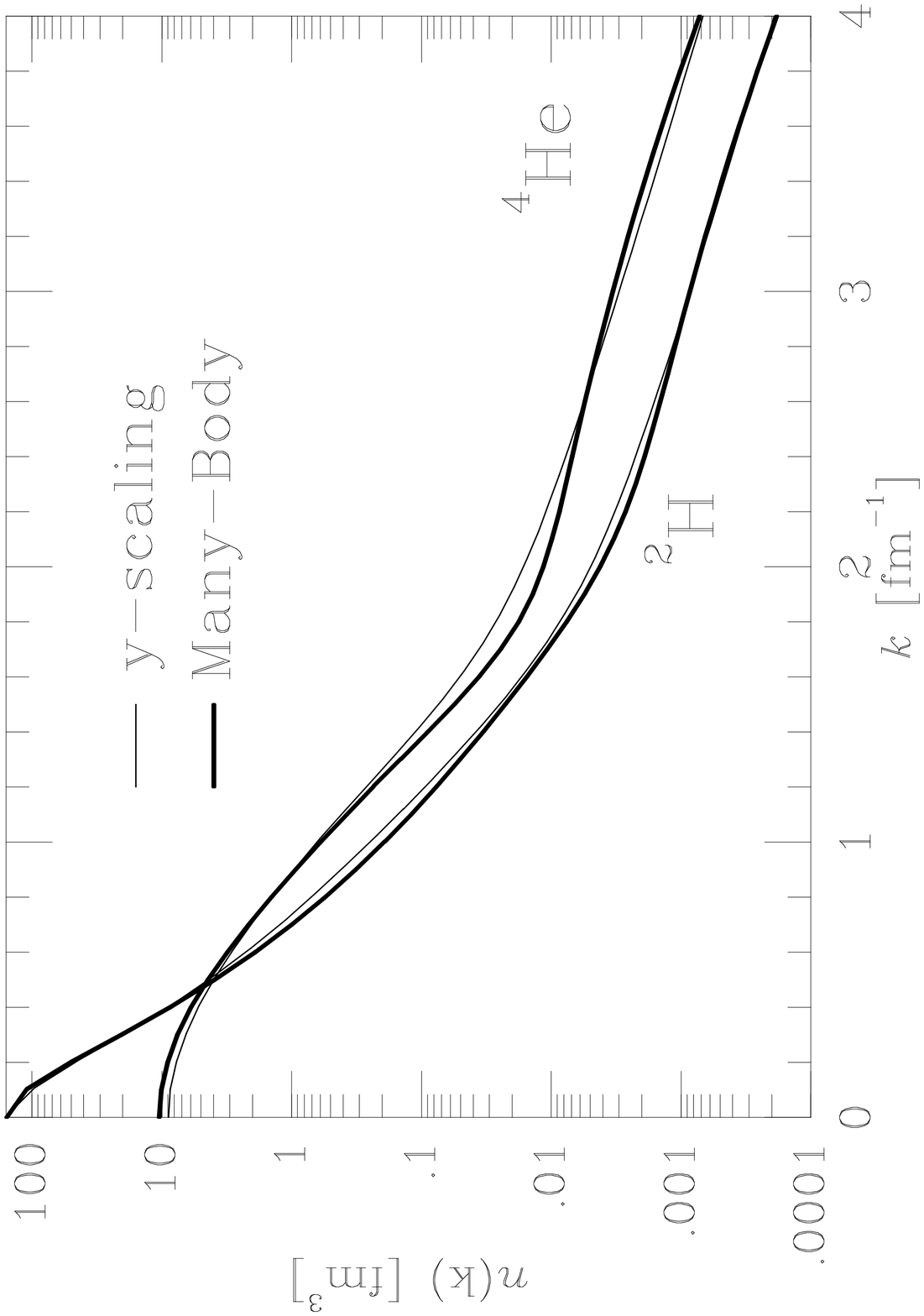}}
\vfill
Fig.~\ref{fig2}.
Ciofi-West Y-scaling

\newpage
\epsfxsize 14cm
\centerline{\epsfbox{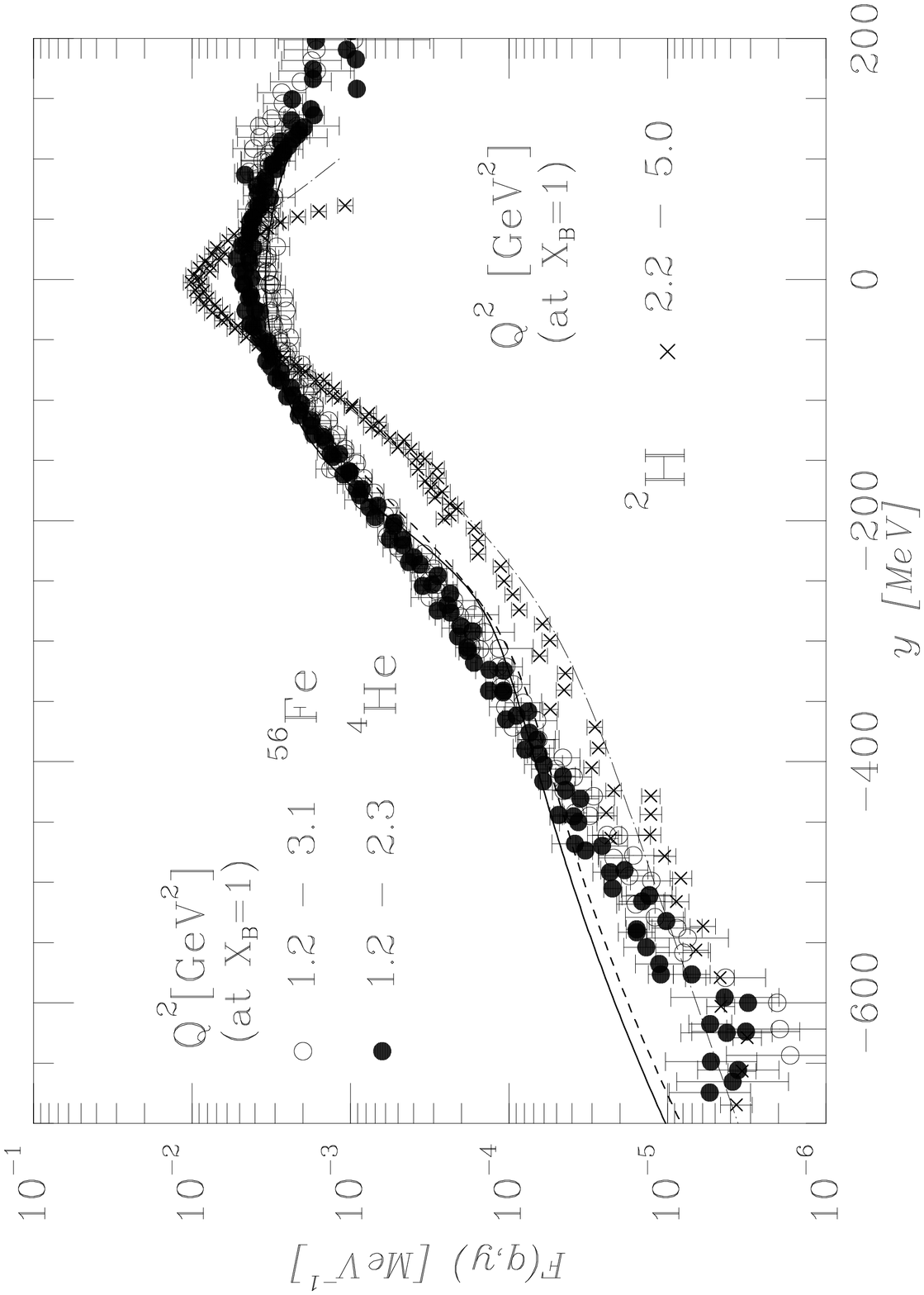}}
\vfill
Fig.~\ref{fig3}.
Ciofi-West Y-scaling

\newpage
\epsfxsize 14cm
\centerline{\epsfbox{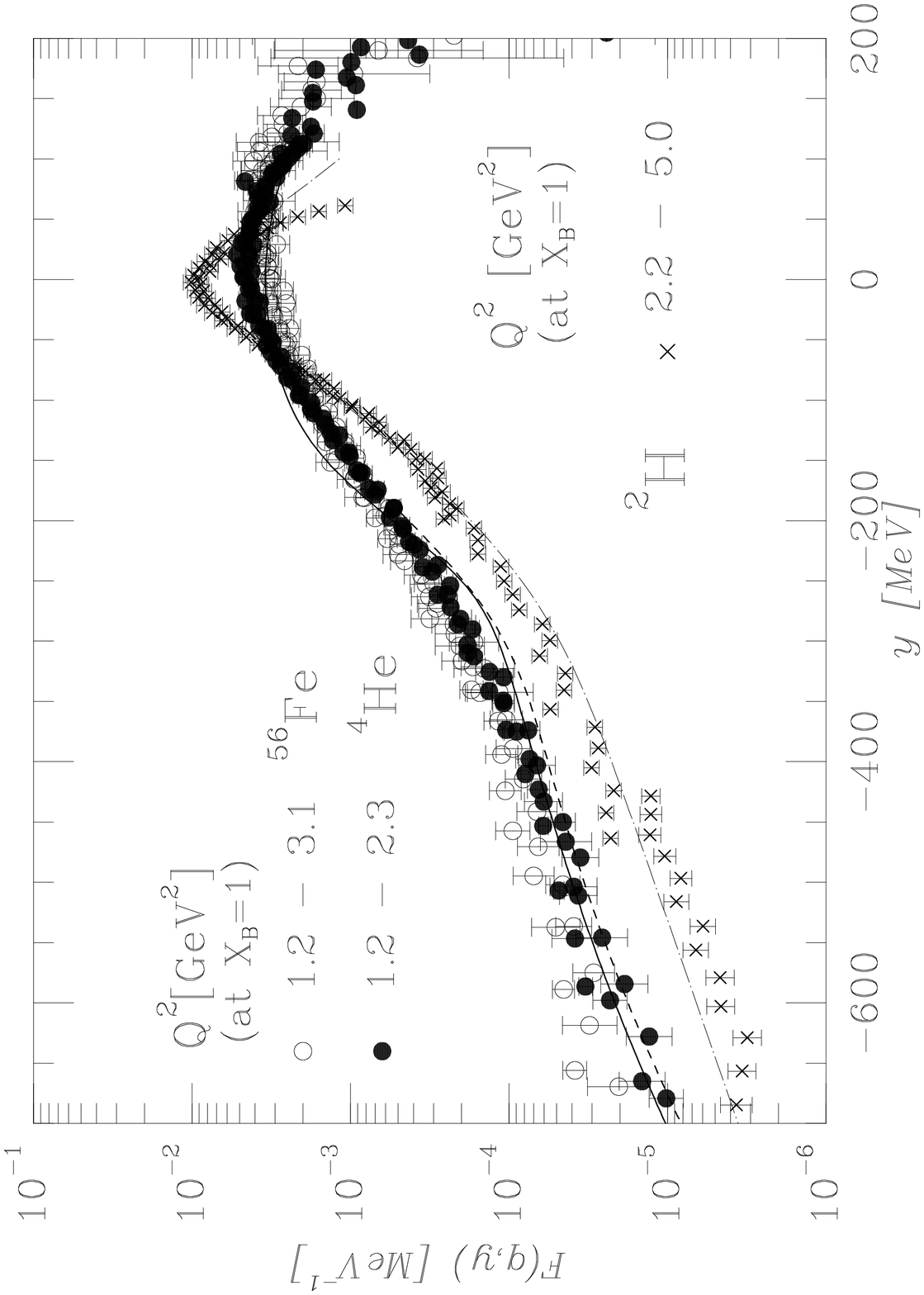}}
\vfill
Fig.~\ref{fig4}.
Ciofi-West Y-scaling


\begin{thebibliography}{99}
\bibitem{foot} Various corrections to this such as meson production,  off-shell effects and so on, will not be considered in
what follows since they do  not play a central role in our discussion.
\bibitem{foot4} Note that  Eq.~(\ref{quattordici}) does not contain free parameters, since $b_A$  and $c_A$ are determined from the Spectral Function and $<E>$ from the Koltun sum rule \cite{clasi}
\end{thebibliography}
\end{document}